\documentclass[twocolumn,twoside,preprintnumbers,amsmath,amssymb,pacs]{revtex4}
\usepackage{epsfig}
\usepackage{graphicx}% Include figure files
\usepackage{dcolumn}% Align table columns on decimal point
\usepackage{bm}% bold math

\def\slashii#1{\setbox0=\hbox{$#1$}             % set a box for #1 
   \dimen0=\wd0                                 % and get its size
   \setbox1=\hbox{\sl/} \dimen1=\wd1            % get size of /
   \ifdim\dimen0>\dimen1                        % #1 is bigger
      \rlap{\hbox to \dimen0{\hfil\sl/\hfil}}   % so center / in box
      #1                                        % and print #1
   \else                                        % / is bigger
      \rlap{\hbox to \dimen1{\hfil$#1$\hfil}}   % so center #1
      \hbox{\sl/}                               % and print /
   \fi}                                         %
\def\slashiii#1{\setbox0=\hbox{$#1$}#1\hskip-\wd0\hbox to\wd0{\hss\sl/\/\hss}}
\def\slashiv#1{#1\llap{\sl/}}
\usepackage{fancyhdr}

\usepackage{pslatex}

\begin{document}

\title{{\Large Fractal Propagators in QED and QCD
and Implications for the Problem of Confinement }}

\author{S. Gulzari, Y. N. Srivastava, J. Swain, A. Widom}

\affiliation{Dept. of Physics, Northeastern University, Boston, MA 02115, USA }

\received{on 15 September, 2006}

\begin{abstract}
We show that QED radiative corrections change the propagator of
a charged Dirac particle so that it acquires a fractional anomalous
exponent connected with the fine structure constant. The result is a
nonlocal object which represents a particle with a roughened trajectory
whose fractal dimension can be calculated. This represents a significant
shift from the traditional Wigner notions of asymptotic states with
sharp well-defined masses. Non-abelian long-range fields are more difficult
to handle, but we are able to calculate the effects due to Newtonian gravitational
corrections. We suggest a new approach to confinement in QCD based on 
a particle trajectory acquiring a fractal dimension which goes to zero
in the infrared as a consequence of self-interaction, representing a
particle which, in the infrared limit, cannot propagate.


\end{abstract}

\maketitle

\thispagestyle{fancy}
\setcounter{page}{0}

\section{Introduction}

In \cite{paper1}, a gauge-invariant calculation was presented showing that
the propagator for a charged particle acquires an interesting fractal 
structure from its self-interaction in quantum electrodynamics. In simple
terms, the usual Dirac propagator $S(k)$ is replaced by the nonlocal expression

\begin{eqnarray}
\nonumber \\ 
S(k)=(\kappa -\slashiv{k}){\cal D}(k),
\nonumber \\
%\tilde{\Delta }(x-y)=
%\int {\cal D}(k)e^{ik\cdot (x-y)}\frac{d^4k}{(2\pi)^4}\ ,
%\nonumber \\ 
{\cal D}(k)=\left(\frac{\kappa}{i\Lambda}\right)^{\alpha /\pi }
\left\{\frac{\Gamma \big(1+(\alpha /\pi )\big)}
{\big[k^2+\kappa^2-i0^+\big]^{\big(1+(\alpha /\pi ) \big)}}\right\},
%\nonumber \\ 
%\tilde{G}(x-y)=\int S(k)e^{ik\cdot (x-y)}\frac{d^4k}{(2\pi)^4}.
\label{GAI27}
\end{eqnarray}
where \begin{math} \hbar \kappa =mc \end{math}, \begin{math} \Lambda  \end{math}
is a short distance length scale, the fractional exponent
\begin{math} \gamma \end{math} is a function of the coupling strength 
\begin{math} \alpha =(e^2/\hbar c) \end{math} which we find to be $\alpha/\pi$
and the usual Gamma function is 
\begin{equation}
\Gamma (z)=\int_0^\infty e^{-s}s^z \frac{ds}{s}
\ \ \ {\rm with}\ \ \ {\Re e}(z) > 0.
\label{AppleG}
\end{equation} 

This represents a rather radical departure from the usual textbook
discussions of particle propagators (with the notable exception of
\cite{JauchandRohrlich}) and from the usual Wigner classification
of elementary particles in terms of a sharp mass and spin (see,
for example, \cite{Weinberg1}). This is despite the fact that it
is well-known that particles coupled to long-range massless fields
cannot have sharp masses\cite{Porrmann,Buchholz1,Buchholz2}. A good
review of the basic issues can be found in \cite{Schroer}. 

Aside from having been neglected in textbooks, the actual exponent
in the above expression has been the subject of some controversy.
There are calculations in the literature based on either infinite sums 
of logarithmic Feynman diagrams\cite{sumlogs} or non-perturbative 
Schwinger\cite{Schwinger,Fradkin,Rivers} computations which argue that such an electron
propagator should be of the {\em form} we obtained:
\begin{equation}
S(k)=\left(\frac{\kappa}{i\Lambda}\right)^{\gamma }
\Gamma (1+\gamma ) \left\{\frac{\kappa -\slashiv{k}}
{\big[k^2+\kappa^2-i0^+\big]^{(1+\gamma )}}\right\},
\label{Appelquist}
\end{equation}.

In the literature there has not been full agreement about what constitutes the 
correct function \begin{math} \gamma (\alpha )\end{math}, nor has there
been a consensus as to whether or not it can be set to zero by a suitable
choice of gauge.

Appelquist and Carazzone\cite{sumlogsAC} argued that 
\begin{math} \gamma =-(\alpha/\pi )+\ldots \end{math} to leading order,
in conflict with earlier work based on summing logarithms\cite{sumlogs}.
Reference \cite{sumlogsAC} does not actually derive the exponent,
rather citing the earlier work, so there is a possibility that a
typographical error may be involved.

The fourth volume of the Landau and Lifschitz course of theoretical
physics\cite{landau} starts off in agreement with
\begin{math} \gamma \neq 0 \end{math} but ultimately sets \begin{math}
\gamma =0 \end{math} by giving a small mass to the photon. Such a photon 
mass explicitly breaks both gauge invariance and conformal/scale invariance
and must be rejected on physical grounds. Intuitively, what goes wrong
is that the fractal structure of the propagator for a charged particle
must be due to the fact that ever increasing wavelengths one can 
produce more and more soft photons without limit. A photon mass
implies a maximum photon wavelength and a loss of scale invariance.

There have been other approaches in the literature, often tackling 
only the case of charged scalar fields, or again arguing that the anomalous
dimension can somehow be set to zero. For example,
a path integral\cite{Rivers} approach  using the
Schwinger\cite{Schwinger} proper time representation of the propagator 
and some work by Bloch and Nordsieck\cite{BN} on soft photon emission
gives the  same sort of result as ours, but with the final answer given only for
charged scalar fields. These results are argued to be gauge invariant in
such a way that the singularity  structure can be returned to a simple
pole by a choice of gauge. As argued above, this is unphysical as there is
a real meaning to a fractional exponent and the failure of charged particles
to have sharp masses. Fried also  discusses this
problem\cite{Friedbook} as do Johnson and Zumino\cite{Johnson_Zumino},
and Stefanis and collaborators\cite{Stefanis}. Batalin, Fradkin and
Schvartsman have made a  similar gauge dependent calculation for scalar
particles\cite{Batalin}.

In addition, there are possible experimental consequences of the
radical change in the nature of the singularity in the propagator.
Handel has argued\cite{Handel} that the
change from  a pole into a branch point has measurable physical
implications for  ``\begin{math} 1/\omega   \end{math}''
noise in the Schr\"{o}dinger (non-relativistic) limit.

The result is clearly not analytic in $\alpha$ and requires a 
non-perturbative approach. Such non-analyticity in $\alpha$
for an all-orders or non-perturbative calculation could be anticipated
on physical grounds from the arguments of Dyson against the convergence
of perturbative expansions in QED\cite{Dyson}. 

Given the apparent confusion in the literature, we present a simple
derivation which preserves gauge invariance throughout. With this
in hand, we then discuss the physical interpretation of the result, extend
it to include Newtonian gravity,discuss the fractal dimension
one can associate with paths involved, and finally introduce a new
way to think about confinement in these terms.

\section{Derivation of the Dressed Electron Propagator}

For completeness, we rederive the expression for the dressed electron
propagator presented in \cite{paper1}.

For an electron in an external electromagnetic field, 
\begin{math} 
F_{\mu \nu}=\partial_\mu A_\nu -\partial_\nu A_\mu ,
\label{GIA1} 
\end{math}
the Dirac propagator 
\begin{eqnarray}
\left(-i\slashiii{d}+\kappa \right)G(x,y;A)=\delta (x-y)\ , 
\nonumber \\ 
d_\mu =\partial_\mu -i\left(\frac{eA_\mu }{\hbar c}\right)\ ,
\label{GIA2} 
\end{eqnarray}
may be solved employing the function 
\begin{math} \Delta (x,y;A) \end{math};  
\begin{eqnarray}
\Delta (x,y;A)=\int \gamma_5G(x,z;A)\gamma_5G(z,y;A)d^4z\ ,
\nonumber \\ 
G(x,y;A)=\left(i\slashiii{d}+\kappa \right)\Delta (x,y;A)\ ,
\nonumber \\  
\left(\slashiii{d}^2+\kappa^2 \right)\Delta (x,y;A) 
=\delta (x-y)\ ,
\nonumber \\
\slashiii{d}^2=-d^\mu d_\mu -
\frac{e}{2\hbar c}\sigma^{\mu \nu }F_{\mu \nu}\ .
\label{GIA3} 
\end{eqnarray}
The Hamiltonian of the electron can be written
\begin{eqnarray}
{\cal H}_{tot}={\cal H}+{\cal H}_{spin}\ ,
\nonumber \\ 
{\cal H}=\frac{1}{2m}\left\{\left(p-\frac{e}{c}A\right)^2+m^2c^2\right\}\ ,
\nonumber \\ 
{\cal H}_{spin}=-\left(\frac{e\hbar }{4mc}\right)\sigma^{\mu \nu }F_{\mu \nu }
\label{GIA4}
\end{eqnarray}
and with \begin{math} p_\mu =-i\hbar \partial_\mu  \end{math}, 
one may define the amplitude for the electron to go from {\it y} to {\it x} in a proper time 
\begin{math} \tau \end{math} as the matrix element 
\begin{equation}
{\cal G}(x,y,\tau ;A)=\left<x\right|e^{-i{\cal H}_{tot}\tau /\hbar }\left|y\right>.
\label{GIA5}
\end{equation}
From Eqs.(\ref{GIA3}), (\ref{GIA4}) and (\ref{GIA5}) follows the electron propagator 
expression 
\begin{eqnarray}
\Delta (x,y;A)=\frac{i\hbar }{2m}\int_0^\infty 
{\cal G}(x,y,\tau ;A) d\tau ,
\nonumber \\ 
\hbar G(x,y;A)=\left(mc-\slashiv{p}+\frac{e}{c}\slashii{A}(x) \right)\Delta (x,y;A)\ .
\label{GIA6}
\end{eqnarray}

One also has a Lagrangian 
\begin{eqnarray}
{\cal L}(v,x;A)=\frac{1}{2}m\left(v^\mu v_\mu -c^2\right)
+\frac{e}{c}v^\mu A_\mu (x),
\nonumber  \\ 
\frac{d}{d\tau}\left(\frac{\partial {\cal L}}{\partial v^\mu}\right)
=\left(\frac{\partial {\cal L}}{\partial x^\mu}\right);
\label{GIA9}
\end{eqnarray} 

As we are interested here in the infrared limit, spin-flips are suppressed
and we can neglect the corresponding term in the Hamiltonian. This corresponds to
usual Bloch-Nordsieck replacement of \begin{math} c\gamma^\mu \end{math} by 
four velocity \begin{math} v^\mu \end{math} -- one simply thinks of the Dirac
spinor having only one nonvanishing component representing an electron of
given spin state and allows no coupling to the other spin state (and, of course,
no coupling to positrons).
 
With this approximation, Eq. (\ref{GIA5}) which gives the propagation
amplitude for an electron can be written in the Lagrangian 
{\em path integral} formulation   
\begin{eqnarray}
{\cal G}(x,y,\tau ;A)\approx \int_{X(0)=y}^{X(\tau )=x}
e^{i{\cal S}[X;A]/\hbar }\prod_\sigma dX(\sigma ), 
\nonumber \\ 
{\cal S}[X;A]=\int_0^\tau {\cal L}(\dot{X}(\sigma ),X(\sigma );A)d\sigma . 
\label{GIA10}
\end{eqnarray}  

This expression deserves some comments. The paths $X(\sigma)$
which are integrated over represent virtual histories for the
electron. The proper time $\tau$ for each path in the sum is not
given by the classical expression
\begin{math} c^2\tau^2 \ne -(x-y)^2 \end{math}, so in integrating
over paths one is integrating here over all proper times.

For each one of these paths, with its own proper time, 
the interaction between the electron and an elecromagnetic vector 
potential $A^\mu$ is described by the action 
\begin{eqnarray}
S_{\rm int}(P;A)=\int_0^\tau {\cal L}_{\rm int}
(\dot{X}(\sigma ),X(\sigma );A)d\sigma ,
\nonumber \\ 
S_{\rm int}(P;A)=\frac{e}{c}\int_0^\tau A_\mu (X(\sigma ))
\dot{X}^\mu (\sigma ) d\sigma ,
\nonumber \\ 
S_{\rm int}(P;A)=\frac{e}{c}\int_P A_\mu (X) dX^\mu ,
\label{GIA11}
\end{eqnarray}
where the integral is along the worldline \begin{math} P \end{math}.

For the case of no external field, we want the action due to self-interaction
(interaction with vacuum fluctuations), so one wants to apply the rule

\begin{eqnarray}
e^{iS_{\rm int}(P;A)/\hbar }\to 
\left<0\right|e^{iS_{\rm int}(P;\hat{A})/\hbar }\left|0\right>_+,
\nonumber \\  
e^{iS_{\rm int}(P;A)/\hbar }\to e^{iS_{self}(P)/\hbar },
\nonumber \\ 
S_{self}(P)=\frac{\hbar \alpha }{2}
\int_P \int_P D_{\mu \nu }(x_1-x_2)dx_1^\mu dx_2^\nu .
\label{GIA12}
\end{eqnarray} 
In the above Eq.(\ref{GIA12}), the subscript 
``\begin{math}+\end{math}'' denotes time ordering, 
\begin{math} \hat{A}_\mu (x) \end{math} denotes the 
operator vector potential field and the photon propagator is given by 
\begin{equation}
D_{\mu \nu }(x_1-x_2)=\frac{i}{\hbar c}
\left<0\right|\hat{A}_\mu (x_1)\hat{A}_\nu(x_2)\left|0\right>_+.
\label{GIA13}
\end{equation}
The action form in Eq.(\ref{GIA12}) is of a well known
form\cite{Feynman-Wheeler}. As noted earlier, we  
have bypassed the usual Bloch-Nordsieck replacement of 
\begin{math} c\gamma^\mu \end{math} by 
four velocity \begin{math} v^\mu \end{math} by simply evaluating a phase 
in the soft photon infrared limit that we are considering here.

The propagator may be written
\begin{eqnarray}
D_{\mu\nu}(x-y)=\left(\eta_{\mu \nu}-
(1-\xi)\frac{\partial_\mu \partial_\nu }{\partial^2}\right)D(x-y),
\nonumber \\ 
D(x-y)=\int \frac{4\pi }{k^2-i0^+}e^{ik\cdot (x-y)}\frac{d^4 k}{(2\pi )^4},
\nonumber \\ 
D(x-y)=\frac{i}{\pi }\left\{\frac{1}{(x-y)^2+i0^+}\right\}.
\label{GIA14}
\end{eqnarray}
where the parameter \begin{math} \xi \end{math} fixes a gauge.

Because the worldline of the electron never begins nor ends 
the partial derivative terms in 
Eq.(\ref{GIA14}) do not contribute to the self action in 
Eq.(\ref{GIA12}). This requirement that the worldline have
no endpoints is required by charge conservation (or, equivalently,
by gauge invariance).

Independent of any choice of the gauge parameter
\begin{math} \xi \end{math} then, we have 
\begin{equation}   
S_{self}(P)=\frac{\hbar \alpha }{2}
\int_P \int_P D(x_1-x_2){dx_1}_\mu dx_2^{\ \mu }.
\label{GIA15}
\end{equation} 

In the absence of any external field (other than that due to 
vacuum fluctuations) we have now 
derived expressions for the renormalized vacuum electron 
propagator
\begin{eqnarray}
\tilde{G}(x-y)=\int S(k)e^{ik\cdot(x-y)}\frac{d^4k}{(2\pi )^4}\ ,
\nonumber \\ 
\tilde{G}(x-y)=\left<0\right|G(x,y;\hat{A})\left|0\right>_+\ ,
\nonumber \\ 
\tilde{G}(x-y)=\left(i\slashiii{\partial }+\kappa \right)
\tilde{\Delta }(x-y)\ ,
\nonumber \\ 
\tilde{\Delta }(x-y)=\frac{i\hbar }{2m}\int_0^\infty 
\tilde{\cal G}(x-y,\tau ) d\tau .
\label{GIA16}
\end{eqnarray}
The functional integral expression for 
\begin{math} \tilde{\cal G}(x-y,\tau ) \end{math}
is given by 
\begin{eqnarray}
\tilde{\cal G}(x-y,\tau )=\int_{X(0)=y}^{X(\tau )=x}
e^{i\tilde {\cal S}[X;A]/\hbar }\prod_\sigma dX(\sigma ), 
\nonumber \\ 
\tilde{\cal S}[X]=\int_0^\tau {\cal L}_0(\dot{X}(\sigma ))d\sigma 
+S_{self}[X] , 
\label{GIA17}
\end{eqnarray}  
wherein the {\em free} electron Lagrangian is 
\begin{equation}
{\cal L}_0(\dot{X})=\frac{1}{2}m_0(\dot{X}^\mu \dot{X}_\mu -c^2),
\label{GIA18}
\end{equation}
and the self action is given by Eqs.(\ref{GIA14}) and (\ref{GIA15}) as 
\begin{equation}
S_{self}[X]=\frac{i\hbar \alpha }{2\pi }
\int_0^\tau \int_0^\tau 
\frac{\dot{X}^\mu (\sigma_1)\dot{X}_\mu (\sigma_2)d\sigma_1 d\sigma_2}
{(X(\sigma_1 )-X(\sigma_2))^2+i0^+}\ .
\label{GIA19}
\end{equation}
The divergent piece of the self action 
\begin{eqnarray}
{\Re }eS_{self}[X]=\frac{\Delta m}{2}\int_0^\tau 
(\dot{X}^\mu (\sigma )\dot{X}_\mu (\sigma )-c^2)d\sigma , 
\nonumber \\ 
|\Delta m|=\infty.
\label{GIA20}
\end{eqnarray}

The formally infinite self-mass can be absorbed into a redefinition of
the finite physical mass \begin{math} 0<m=(m_0+\Delta m)<\infty \end{math}. 
Thus, Eq.(\ref{GIA17}) is renormalized to 
\begin{eqnarray}
\tilde{\cal G}(x-y,\tau )=\int_{X(0)=y}^{X(\tau )=x}
e^{i\tilde {\cal S}[X;A]/\hbar }\prod_\sigma dX(\sigma ), 
\nonumber \\ 
\tilde{\cal S}[X]=\int_0^\tau {\cal L}_m(\dot{X}(\sigma ))d\sigma 
+iW[X]
\nonumber \\  
{\cal L}_m(\dot{X})=\frac{1}{2}m(\dot{X}^\mu \dot{X}_\mu -c^2)
\nonumber \\  
W[X;\tau ]={\Im }mS_{self}[X],
\nonumber \\  
W[X;\tau ]=\frac{\hbar \alpha }{2\pi }
\int_0^\tau \int_0^\tau 
\frac{\dot{X}^\mu (\sigma_1)\dot{X}_\mu (\sigma_2)d\sigma_1 d\sigma_2}
{(X(\sigma_1 )-X(\sigma_2))^2}\ .
\label{GIA21}
\end{eqnarray}  
For a straight-line path \begin{math} X^\mu (\sigma )=V^\mu \sigma  \end{math} 
with \begin{math} V^\mu V_\mu =-c^2  \end{math}, one finds 
\begin{equation}
W(\tau )=\frac{\hbar \alpha }{2\pi }
\int_0^\tau \int_0^\tau \frac{d\sigma_1d\sigma_2}{(\sigma_1-\sigma_2)^2}.
\label{GIA22}
\end{equation}
This expression as it stands is infinite and requires regularization. 
Using differential regularization\cite{Freedman} we have
\begin{equation}
\frac{d^2W(\tau )}{d\tau ^2}=\frac{\hbar \alpha }{\pi \tau^2}\ .
\label{GIA23}
\end{equation}
The solution to the differential equation Eq.(\ref{GIA23}) with a logarithmic cut-off 
\begin{math} \Lambda  \end{math} is 
\begin{equation}
W(\tau )=-\left(\frac{\hbar \alpha }{\pi }\right)\ln \left(\frac{c\tau }{2\Lambda }\right).
\label{GIA24}
\end{equation}
From Eqs.(\ref{GIA21}) and (\ref{GIA24}), one finds 
\begin{equation}
\tilde{\cal G}(x-y,\tau )\approx e^{-W(\tau )/\hbar }
\tilde{\cal G}_m(x-y,\tau )
\label{GIA25}
\end{equation}
wherein \begin{math} \tilde{\cal G}_m(x-y,\tau ) \end{math} is the proper 
time Green's function for a particle of mass \begin{math} m \end{math}
with the corresponding free Lagrangian \begin{math} {\cal L}_m(\dot{X}) \end{math}. 
To exponentially lowest order in \begin{math} \alpha  \end{math}, one
then has
\begin{eqnarray}
\tilde{\cal G}_m(x-y,\tau )=\int \left\{e^{-i\hbar (k^2+\kappa^2)\tau /2m}
e^{ik\cdot(x-y)} \right\}\frac{d^4k}{(2\pi )^4}\ ,
\nonumber \\ 
\tilde{\cal G}(x-y,\tau )=\left(\frac{c\tau }{2\Lambda}\right)^{\alpha /\pi }
\tilde{\cal G}_m(x-y,\tau ).\ \ 
\label{GAI26}
\end{eqnarray}
Eqs.(\ref{GIA16}) and (\ref{GAI26}) then imply 
\begin{eqnarray}
\nonumber
\tilde{\Delta }(x-y)=
\int {\cal D}(k)e^{ik\cdot (x-y)}\frac{d^4k}{(2\pi)^4}\ ,
\nonumber \\ 
\tilde{G}(x-y)=\int S(k)e^{ik\cdot (x-y)}\frac{d^4k}{(2\pi)^4}.
\label{GAI27-2terms}
\end{eqnarray}
and we obtain Eq. (\ref{GAI27}).

We chose differential regularization as simple and convenient, and
preserving gauge invariance, but other regularizations can also be
used as long as they are also gauge invariant. We plan to return to this
issue in a later publication in more detail. In a sense, the form
of the answer is intuitively clear since, on dimensional grounds,
the expression to be regularized can only be logarithmic. The sign, as
we shall see below, has a physically sensible interpretation, and is
in agreement with reference \cite{JauchandRohrlich} and with reference
\cite{sumlogsAC} if one assumes that there was a typographical error
going from reference \cite{sumlogs}. 

\section{A Physical Interpretation}

It is interesting to consider what the physical interpretation of the non-integer
exponent in the radiatively corrected Dirac propagator means. First of all, the fact
that the exponent is non-integer means that the renormalized Dirac operator is  
non-local\cite{fraccalc}. This was, of course, to be expected since the 
electromagnetic field has infinite range.

Non-locality has been 
previously introduced ad-hoc\cite{nonlocalreg,Moffat} as a regularization tool. 
Here it appears naturally, suggesting that some form of regularization of
otherwise formally divergent expressions may be implicit in at least some
quantum field theories with massless fields, but appearing only when one
goes beyond perturbation theory. We will also see later that there is 
evidence that the corrections to the exponent can be changed when other fields
are added, and need not always be of the same sign as in pure QED.
The appearance of this sort of non-locality also makes the analytic but
rarely used regularization proposed by Speer\cite{Speer} more physically
motivated.

So far we have only been able to treat the long-wavelength approximation
since we expect a quenched approximation in which we can ignore electron
loops to be a reasonable one. The physical picture would be one where,
as one backs away from the worldine of an electron, one continues to 
see photons radiated and absorbed, but now of longer and longer wavelength.
This would suggest a fractal\cite{Mandelbrot} 
structure, which is made precise by the above derivation. Such notions of scaling 
and fractality are not new in QED and in quantum field theory in general, but are 
often considered in the high energy, ultraviolet limit\cite{Collins,GL}.
This continues without limit at longer and longer distance scales
since there is no minimum energy photon (photons are massless) and this
self-similarity reflects the scale invariance of Maxwell's equations.

If the photon is given a mass, however small, this structure will break down 
asymptotically, since now there is a minimum energy required to create a virtual 
photon, and at distances greater than the corresponding Compton wavelength one will 
get the non-interacting Dirac propagator. This was by
Lifshitz {\em et al.}\cite{landau}, and this argument makes clear how breaking gauge 
invariance, {\em i.e.} including a photon mass, removes the anomalous scaling behavior 
here derived for gauge invariant QED. 

The fact that a particle is non-localizable, at least in part due to its 
electromagnetic field which extends over all space, is interesting. The feeling  
of this calculation is such that at least part of what one thinks 
of as quantum-mechanical about an otherwise point-particle (its lack of 
localizability) may arise from the non-perturbative quantum mechanics of its 
self-interaction\cite{Hestenes}.

At shorter distance scales one might again expect an anomalous dimension, but
now the calculations are more complicated since one must imagine more and
more electron (and other charged particle) loops contributing, with an 
ever ``frothier'' structure at smaller and smaller distance scales. In fact,
one would not even expect a constant exponent which is independent of momentum
since one expects a running $\alpha$ with values which change with momentum
scale. In the infrared limit one simply goes to $\alpha(q^2=0)$ which is
the Thompson value and there are no additional complications. In the ultraviolet
limit, one expects a continuously changing exponent and thus a multifractal
as opposed to fractal structure. In addition, the lack of asymptotic freedom
leads one to expect trouble at very short distance scales unless, as noted
above, other fields enter significantly.

\section{Newtonian Quantum Gravity}
\label{sec:gravity}

For quantum gravity one can do the analysis in much the same way as for 
quantum electrodynamics. The ultraviolet divergences need not worry us  
since we are dealing with a strictly infrared problem. There should be 
graviton-graviton interactions, but if we neglect these as 
small compared to graviton-electron interactions we can just repeat what was done for 
electrostatics but now with the Newtonian limit of gravity. 

This approximation would not be as reasonable in the QCD case for a quark
propagator since gluon-gluon couplings 
are comparable to gluon-quark couplings, but there is evidence from other
calculations of the appearance of anomalous dimensions for infrared propagators.
(For a review see \cite{Alkofer}. Very recent calculations 
can also be found in references \cite{Fischer}.)

Perhaps the simplest way to think of this is to look at Eq. (\ref{GIA15}) and
regard the (singular) double integral as the limit of two paths which
must be taken as approaching each other arbitrarily closely. This is an
alternative to the differential regularization we used in Eq.(\ref{GIA23})
and physically corresponds to two electron worldlines interacting via
the exchange of an arbitary number of photons between them and at all
points of their paths. In the limit of the paths coinciding, this
becomes self-interaction, with the electron continuously exchanging
photons with itself along its worldline.

For gravity (at least in the Newtonian limit) we want to replace the
repulsive electrostatic self-interaction,  
say \begin{math} +(e^2/r)  \end{math}, with the attractive 
gravitational self-interaction, say \begin{math} -Gm^2/r  \end{math},
suggesting an asymptotic form of the Dirac propagator exponent    
\begin{equation}
\gamma \approx \frac{1}{\pi \hbar c }(e^2-Gm^2)+\ldots \ . 
\end{equation}
If \begin{math} m=|e/ \sqrt{G}|\end{math}, which is the ADM\cite{Ashtekar,ADM} 
mass of charged shell of charge \begin{math} e \end{math}, regularized by its own gravity, then one 
recovers an effectively free propagator. Since one generally 
makes measurements using the electromagnetic interaction, and the suggestion that 
quantum mechanics might be linked to self-interaction\cite{Hestenes}, it is interesting 
to consider what this might imply for the role of gravity in the quantum measurement 
problem\cite{Penrose}. In particular, since one has a connection between mass 
and charge which is non-perturbative in Newton's \begin{math} G \end{math}, and implies 
a mass near the Planck mass \begin{math} \sim 10^{-5} \end{math} gm, which may be thought 
to be in the neighborhood of a putative classical-quantum boundary.

Intuitively, electromagnetic interactions roughen the path of an electron, making
it ``spread out more'', which is natural since the electromagnetic force between
two identical charges (or an electron and itself) is repulsive. Gravitational
interactions, on the other hand,
make it ``spread out less'', since the gravitational force between two identical
objects (here an electron and itself) is attractive.

\section{Fractal Paths}

The notion of a path somehow ``spreading out'' can be made
precise with the notion of fractal dimensions.

The fractal nature of particle
paths has been discussed in the literature (see, for
example \cite{FeynmanHibbs,HeyFractals}). 
Abbott and Wise\cite{AbbotandWise}, working within nonrelativistic quantum
mechanics, found 2 as the dimension of a
quantum mechanical path, as opposed to 1 for a classical path.
Cannata and Ferrari\cite{Cannata}
extended this work for spin-1/2 particles and find different results not only
in the classical and quantum mechanical limits, but also in the
non-relativistic and relativistic limits.

Intuitively one can understand the dimension 2 result for the nonrelativistic
quantum mechanical case by
thinking of the Schr\"{o}dinger equation as a diffusion equation in imaginary
time\cite{Nelson}. For diffusion one has the distance $r$ a particle covers in time $t$
satisfying a relationship of the form $t\propto r^d$ wherein $d$ is the fractal 
dimension of the ``path''. For example, $t\propto r^2$ in the diffusion limit,
and $t\propto r$ in the ballistic (simple path) limit\cite{Sornette}. 

Here we have a closely analogous situation but with a 4-dimensional Hamiltonian 
${\cal H}$  and with fractal diffusion in proper time, and as was showed in
\cite{paper1}, one has

\begin{equation}
d=2(1+\gamma )\approx 2+\frac{2\alpha }{\pi }+\ldots \ ,
\label{PI1}
\end{equation}

All previous discussions of fractal propagators in quantum field theory and quantum
mechanics have, to the best of our knowledge, ignored the effects of
self-interaction via long-range fields.

The fact that self-interaction can qualitatively change the nature
of a propagating particle even when coupled to a weak (small coupling
constant) Abelian gauge field naturally leads one to wonder whether or
not coupling to a strong (large coupling constant) non-Abelian gauge
field might have significant physical effects. We turn now to a discussion
of possible implications of this sort of phenomenon for an understanding
of confinement in QCD.

\section{QCD}

In this section we consider how propagators
with anomalous dimensions could shed light on the confinement
mechanism in QCD. Precise and rigorous calculations are beyond
the scope of this paper, but we do indicate approaches to such 
calculations and what one would expect qualitatively. 

Quantum chromodynamics is an incredibly difficult theory in which to say
anything precise. The fundamental degrees of freedom carry strong (colour)
charges, interact via highly nonlinear interactions, and (by hypothesis),
are not even observable asymptotically.

Perturbative treatments make QCD look very much like QED, and involve
propagators which look pretty much like those of QED with the main
difference being in the nonabelian nature of the gauge fields and their
associated self-couplings.
The question which we now
raise is whether or not one can imagine a situation whereby anomalous
exponents in the infrared will give rise to such strong qualitative changes
in how propagators behave that the originally postulated degrees of freedom 
will simply not propagate (at least at low energies) -- that is, that some
form of confinement appears.

What could confinement mean in terms of an anomalous dimension? If in
Eq. \ref{PI1} we had a $\gamma$ which drove $d$ to zero, one would
have a zero-dimensional path, which is no path at all! This would
correspond to a confined particle.

Let us see how far we can argue that such a phenomenon would occur.
One could start from Eq.(\ref{GIA15}) with an appropriate form of $D(x_1-x_2)$
for QCD valid for any separation. Unfortunately, we do not have an
exact expression for $D(x_1-x_2)$
in the infrared limit and in fact every expectation is that (even aside
from colour indices) it is very 
different from the QED one -- presumably having a confining term, roughly
linear in separation. Of course this would be {\it assuming} confinement.

A simpler approach is to imagine that $d=2(1+\gamma)$ where $\gamma\propto\alpha_s$
and $\gamma<0$. If the strong coupling constant $\alpha_s$ is large enough then
one could get $d\rightarrow 0$. Let us follow this line of argument more
closely. The first thing to check is that the energy of self-interaction of
a quark with itself is indeed of the same sign as that due to gravity (and
opposite to that due to electromagnetism). While the interaction energy between
a quark  and an antiquark in the singlet state separated by a distance $r$ 
is given perturbatively
by $\frac{4}{3}\frac{\alpha_s(r)}{r}$, the corresponding quark-quark expression
in the triplet is $-\frac{2}{3}\frac{\alpha_s(r)}{r}$. In other words, it is
of the correct sign to lead to confinement.

Now let us revisit  Eq.(\ref{GIA15}) and
regard the (singular) double integral as the limit of two paths which
must be taken as approaching each other arbitrarily closely as we did in section
\ref{sec:gravity}. We are now looking at two paths of a red, say, quark,
exchanging an arbitrary number of gluons. If we now argue as we did for
gravity, we have $d=2(1-\frac{2}{3}\frac{\alpha_s(r)}{\pi})$. It is
well-known\cite{Yogi1,Yogi2}
(and can be shown on very general grounds from dispersion relations)
that $\alpha_s$ must increase without bound as $r$ goes to infinity (or
squared momentum transfer $q^2$ goes to zero). Happily
here one only needs $\alpha_s$ to reach a value making $d=0$. At the
corresponding value of $q^2$ the dimension of a quark path goes to zero
and the quark is confined. Lower values $q^2$ (and negative dimensions)
make no sense since the integrals in Eq. (\ref{GIA15}) now have no paths
any more as soon as $d$ hits zero. Note that
this approach to thinking about confinement does not require any singular
(i.e. infinite) values of $q^2$ (or length) but should happen at some well-defined
value of $q^2$ (or length) presumably related to the confinement scale.

A rigorous calculation along these lines would be difficult, but a few
points can be made in defense of this overall picture. First of all, why
would one imagine in the regularized view of Eq. (\ref{GIA15}) that
one was always looking at a red quark interacting
with another red quark? Surely each emitted gluon changes the colour of
the particle and complicates matters. This is indeed true, but if one
has defined the quark as being red (in some gauge) then {\em on average}, even
with various colours of gluons being continuously emitted, one expects
that it would remain {\em on average} red. Any tendency to be green or blue would be
a violation of $SU(3)$ symmetry since it would indicate a preferred direction
in colour space other than ``red'' (The effective value of the colour charge
would be excpected to be scale-dependent, but this is exactly what a running
coupling constant is meant to describe.) This sort of argument could be the
basis for a more rigourous mean field theory approach. Second
$\alpha_s$ is usually studied as a function of $q^2$, but its form
is not even known at long distance scales and thus
a rather difficult (!) function to work out in coordinate space to use
in Eq.(\ref{GIA15}). The point we want to make here is that it {\it is} 
very generally known that $\alpha$ increases in the infrared limit, and 
this alone is enough to argue for confinement. Even without an explicitly
known confining form for $D(x-y)$, all one needs to know is that $\alpha$
continues to rise. At some point it will rise to a value high enough to
drive $d$ to zero. Beyond that, the quark is non-propagating (our version of
``confined'') and there is no need
to evalute paths in Eq.(\ref{GIA15}) since the quark has none! The expected
results from asymptotic freedom are, of course, reproduced, since as $\alpha_s$
goes to zero the propagator goes over to its free form (aside from whatever
corrections remain due to charge and mass as discussed earlier -- particles
coupled to long-range fields never really get free!).

A similar argument would be expected to hold for any coloured objects,
including gluons and hypothetical objects in other representations of
$SU(3)$ colour. 

The physical picture is an interesting one and rather different from
the assumption that quarks are bound due to strong attractive forces
to other quarks which increase linearly with separation in some approximation.
Here, below some energy scale (that at which the dimension $d$ goes to zero),
coloured objects interact so much with their own glue that they get ``stuck''
in the sense that their would-be paths are reduced to dimension zero. At 
higher energy scales they partially escape, with a dimension which rises
as the effective coupling strength drops until they become, to a good
approximation, free. It is interesting to note that the view of confinement
suggested here applies to any coloured object with no need for it to have
neighbours present to ``bind'' to it and ensure that a colour singlet
state propagates -- here interaction of a single coloured object
with its own colour field is enough to confine it.

\section{Conclusions}

We have reviewed the simple and intuitive path integral description of how the 
propagator for a charged Dirac particle is modified by soft self-energy radiative 
corrections as shown in reference \cite{paper1}. 
The result is a self-similar (fractal) object with the non-locality one would 
expect for a particle carrying an infinite range field. Arguments are made
for a similar, but qualitatively different, effect due to attractive self-interactions
such as gravity and a calculation made for Newtonian gravity. The results are
linked to the fractal dimensions of the paths that particles take in quantum
field theory, and the effects of repulsive (QED) and attractive (gravity)
self-interactions are discussed.
Finally an attempt is made to estimate what effects would be
expected in QCD, with a link made to confinement at a finite energy scale
in terms of a fractal dimension which goes to zero.

\section{Acknowledgements}

We would like to thank the organizers of the IRQCD 2006 conference for a
wonderful and stimulating
conference where one of us (J.S.) presented an earlier version of this work.
We would also like to thank Reinhard Alkofer for pointing out
references \cite{JauchandRohrlich}
and \cite{Alkofer} to us, and Christian Fischer for pointing out \cite{Fischer}.
This work was supported in part by the National Science Foundation under grant
NSF-0457001.

\end{document}